\documentstyle[prl,epsfig,aps,multicol]{revtex}  
\title{Renormalisation in Quantum Mechanics}

\author{ 
H.~Jirari$^{a}$,
H.~Kr\"{o}ger$^{a}\footnote{Corresponding author, Email: hkroger@phy.ulaval.ca}$, 
X.Q.~Luo$^{b}$, 
G.~Melkonyan$^{a}$, 
and K.J.M.~Moriarty$^{c}$
 }

\address{$^{a}$D\'{e}partement de Physique, Universit\'{e} Laval, 
  Qu\'{e}bec, Qu\'{e}bec G1K 7P4, Canada \\ 
  $^{b}$Department of Physics, Zhongshan University, 
  Guangzhou 510275, China \\
  $^{c}$Department of Mathematics, Statistics and Computer Science,
  Dalhousie University, Halifax, Nova Scotia B3H 3J5, Canada }
      
\date{\today} 

\begin{document} 

\newcommand{\co}[1]{{\bf[#1]}}
\newcommand{\be}{\begin{equation}}
\newcommand{\ee}{\end{equation}}
\newcommand{\bea}{\begin{eqnarray}}
\newcommand{\eea}{\end{eqnarray}}
\newcommand{\sgn}{\mbox{sgn}}
\newcommand{\slesssim}{{\scriptstyle \lesssim}}
\newcommand{\ewxy}[2]{\setlength{\epsfxsize}{#2}\epsfbox[10 60  640 770]{#1}}

\renewcommand{\vec}[1]{{\bf #1}}

\maketitle

\begin{abstract}
We study a recently proposed quantum action depending on temperature. We construct a renormalisation group equation describing the flow of action parameters with temperature. At zero temperature the quantum action is obtained analytically and is found free of higher time derivatives. It makes the quantum action an ideal tool to investigate quantum chaos and quantum instantons.
\end{abstract}


\begin{multicols}{2}

{\it 1. Introduction}.-- Renormalisation is a well known concept in $QFT$ 
as well as in Q.M. \cite{QM_Renorm}. In Ref.\cite{Jirari1} we proposed the conjecture that to each classical action 
$S = \int dt \frac{m}{2} \dot{x}^{2} - V(x)$ 
corresponds a quantum/renormalized action 
$\tilde{S} = \int dt \frac{\tilde{m}}{2} \dot{x}^{2} - \tilde{V}(x)$.
The quantum action takes into account quantum effects via renormalized action parameters.
The physical principle is well known from solid state physics: 
A charged particle propagating in a solid interacts 
with atoms. This results in an effective mass and charge, different from that of free propagation. Here we refer to renormalisation to describe the difference in propagation between classical and quantum physics. The definition of the quantum action $\tilde{S}$ is 
\be
\label{DefRenormAction}
G(x_{fi},t_{fi}; x_{in},t_{in}) = \tilde{Z} 
\exp [ \frac{i}{\hbar} \left. \tilde{S}[\tilde{x}_{cl}] 
\right|_{x_{in},t_{in}}^{x_{fi},t_{fi}} ] ~ ,
\ee
where $G$ is the transition amplitude, $\tilde{x}_ {cl}$ is the classical path, which minimizes the action $\tilde{S}$ and 
$\tilde{Z}$ is a normalisation factor. 
Eq.(\ref{DefRenormAction}) is valid with 
the {\em same} $\tilde{S}$ and $\tilde{Z}$ for all sets of 
boundary positions $x_{fi}$, $x_{in}$ for a given time interval $T=t_{fi}-t_{in}$. Any dependence on $x_{fi}, x_{in}$ enters via the trajectory 
$\tilde{x}_ {cl}$. $\tilde{Z}$ and the parameters of $\tilde{S}$ depend on time $T$.

It has become very popular to study classical and quantum chaos in billard systems. Milner et al.\cite{Milner:01} and Friedman et al.\cite{Friedman:01} 
studied the motion of ultra-cold atoms in a billard formed by laser beams. 
Dembrowski et al.\cite{Dembrowski:01} tested the prediction of a generalized semi-classical trace formula for billards in the regime between regular and chaotic motion. The quantum action is a new link between classical and quantum physics. In Refs.\cite{Jirari2,WCNA} the quantum action has been used to define quantum instantons and quantum chaos, and to present first numerical results.  
Here we ask:
(A) Can one obtain the quantum action analytically when the temperature goes to zero or infinity? 
(B) Is the quantum action suitable to describe a double well potential with a bi-stable ground state?
(C) In $QFT$ the renormalisation group equation predicts the flow of parameters when changing the scale. Can one find a similar equation for the quantum action when changing temperature?

{\it 2. Analytic quantum action for small and large T}.-- In statistical field theory with a single phase transition, there are two fixed points of the renormalisation group, namely at zero and infinite temperature. In those limits we find an analytical form for the quantum action. 
(i) Limit $T \to 0$. In this limit, according to Dirac, 
the transition amplitude is given by the classical action,
\be
\label{DiracAction}
G(x,T;x_{in},0) \stackrel{T \to 0}{\longrightarrow} Z 
\exp[ -  S[x_{cl}]|_{x_{in},0}^{x,T} / \hbar  ] ~ . 
\ee
Consequently, we have the analytical result $\tilde{Z} = Z$, $\tilde{S} = S$.
(ii) Limit $T \to \infty$. In this limit the transition amplitude is determined by the ground state (Feynman-Kac),
\be
\label{FeynmanKac}
G(x_{fi},T;x_{in},0) \stackrel{T \to \infty}{\longrightarrow} 
\langle x_{fi}|\psi_{gr} \rangle e^{-E_{gr} T/ \hbar} 
\langle \psi_{gr}|x_{in} \rangle ~ .
\ee
We assume that the quantum potential $\tilde{V}(x)$ is a positive, parity symmetric function increasing without bounds when $|x| \to \infty$ ("confining potential"), with a non-degenerate minimum at $x=0$. As shown in Fig.[1],
the trajectory $\tilde{x}_{cl}$ starts at $x_{in}$ and arrives at $x_{fi}$ such that the action $\tilde{S} = \int_{0}^{T} dt ~ \tilde{m} \dot{x}^{2}/2 + \tilde{V}(x)$ becomes a minimal.  
For large $T$ the trajectory stays most of the time close to the bottom of the potential valley $\tilde{V}=\tilde{v}_{0}$.
Energy is conserved,
$\epsilon = - \tilde{T}_{kin} + \tilde{V} = \mbox{const}$.
In the limit $T \to \infty$, this implies 
$\tilde{V} \to \tilde{v}_{0}$, $\tilde{T}_{kin} \to 0$ 
and $\epsilon \to \tilde{v}_{0}$ (note that $\tilde{v}_{0}$ is independent of $x_{in}$, $x_{fi}$).
Then holds
\bea
\label{ActFeynKac}
&& \tilde{\Sigma} \equiv \tilde{S}[\tilde{x}_{cl}]|_{x_{in},0}^{x_{fi},T} 
= \int_{0}^{T} dt ~ \tilde{T}_{kin} + \tilde{V} 
= \int_{0}^{T} dt ~ 2 \tilde{T}_{kin} + \epsilon
\nonumber \\
&& = \tilde{v}_{0} T + \tilde{m} \int_{0}^{T} dt ~ \dot{x}^{2}
= \tilde{v}_{0} T + \tilde{m} \left( \int_{0}^{T/2} 
+ \int_{T/2}^{T} dt ~ \dot{x}^{2} \right)
\nonumber \\
&& = \tilde{v}_{0} T + \tilde{m} \left( 
\int_{x_{in}}^{0} + \int_{0}^{x_{fi}} dx ~ \dot{x} \right)
\nonumber \\
&& = \tilde{v}_{0} T + \left( 
\int_{x_{in}}^{0} + \int_{0}^{x_{fi}} dx ~  
_{\pm} \sqrt{2 \tilde{m}(\tilde{V}(x) -\tilde{v}_{0}) } 
\right) .
\eea
Energy conservation allows to express $\dot{x} = \pm \sqrt{ \frac{2}{\tilde{m}} (\tilde{V}(x) - \tilde{v}_{0}) }$, 
where the sign depends on initial and final data. In Fig.[1] we considered the case: $x_{in} < 0$, $\dot{x}_{in} > 0$, 
$x_{fi} >0$, $\dot{x}_{fi} > 0$. The conjecture says
$G = \tilde{Z} \exp[- \tilde{\Sigma}/\hbar]$. Subsuming 
$\tilde{Z}/\tilde{Z}_{0}$ into $\tilde{\Sigma}$ (where $\tilde{Z}_{0}$ is a constant of dimension $1/L$) this gives
\bea
\label{TransAmplQuantPot}
&& G(x_{fi},T;x_{in},0) = \tilde{Z}_{0} ~ e^{-\tilde{v}_{0} T/\hbar} ~ \times
\nonumber \\
&& e^{ - \int_{0}^{x_{fi}} dx ~ 
\sqrt{2 \tilde{m}( \tilde{V}(x) -\tilde{v}_{0} ) }/\hbar } 
~ e^{ - \int_{x_{in}}^{0} dx ~ 
\sqrt{2 \tilde{m}( \tilde{V}(x) - \tilde{v}_{0} ) }/\hbar } . 
\eea
Comparison of Eqs.(\ref{FeynmanKac},\ref{TransAmplQuantPot}) shows agreement in the form.
The wave function normalisation misses subleading terms in the $1/T$ expansion of $\tilde{v}_{0}$ \cite{Jirari1},  
$\tilde{v}_{0}(T) \sim_{T \to \infty} A + B/T + C/T^{2} + \cdots$.
Including the term $B$ gives the proper normalisation.
Comparing Eqs.(\ref{FeynmanKac},\ref{TransAmplQuantPot}) yields 
\be
\label{GroundStateLaw}
\psi_{gr}(x) = \frac{1}{N} ~ e^{ - \int_{0}^{|x|} dx' ~ 
\sqrt{2 \tilde{m}( \tilde{V}(x') - \tilde{v}_{0} ) }/\hbar }, ~
E_{gr} = \tilde{v}_{0} ,
\ee
with $N = (\tilde{Z}_{0} ~ \exp[-B/\hbar])^{-1/2}$. From this we compute 
\bea
\hbar^{2} \frac{\psi_{gr}''(x)}{\psi_{gr}(x)} = 2 \tilde{m} 
( \tilde{V}(x) - \tilde{v}_{0} ) 
 - \frac{ \hbar \tilde{m} ~ \frac{d}{dx} \tilde{V}(x) }
{ \sqrt{2 \tilde{m}( \tilde{V}(x) - \tilde{v}_{0} ) } }
\mbox{sgn}(x) ~ .
\eea
Comparison with the stationary Schr\"odinger equation
\be
\label{StatSchrodEq}
\hbar^{2} \frac{ \psi_{gr}''(x) }{ \psi_{gr}(x) } = 2m ( V(x) - E_{gr}) ~ ,
\ee
leads to the transformation law
\bea
\label{TransformLaw}
&& 2 m(V(x) - E_{gr}) =  
\nonumber \\
&& 2 \tilde{m}(\tilde{V}(x) - \tilde{v}_{0}) 
- \frac{\hbar}{2} \frac{ \frac{d}{dx} 2 \tilde{m} (\tilde{V}(x) - \tilde{v}_{0})}
{ \sqrt{2 \tilde{m}( \tilde{V}(x) - \tilde{v}_{0} ) } } \mbox{sgn}(x) ~ .
\eea
Example. Consider a particle of mass $m$ moving in the potential 
$V(x)=v_{4} x^{4} -\hbar \sqrt{2v_{4}/m} |x|$. 
Choosing $\tilde{m}=m$, Eq.(\ref{TransformLaw}) determines the quantum potential
and Eq.(\ref{GroundStateLaw}) reproduces exactly the ground state wave function $\psi_{gr}(x) = \frac{1}{N} \exp[- \sqrt{2m v_{4}}|x|^{3}/(3\hbar)]$
and energy $E_{gr} = 0$.

Another example is the inverse square potential, 
given by the potential $V(x) = \frac{1}{2} m \omega^{2} x^{2} + g/x^{2}$, 
$g>0$. It can be considered as a double well potential with an infinitely high wall at the origin. We consider the motion of the particle only for $x > 0$.
The transition amplitude is known analytically \cite{Schulman}. 
The ground state energy is given by
$E_{gr}=\hbar \omega (1 + \gamma)$, with $\gamma = \frac{1}{2} [1 + 8 m g/\hbar^{2} ]^{1/2}$. Its wave function is $\psi_{gr}(x) = N^{-1} x^{\gamma + 1/2} \exp[- \frac{m \omega}{2 \hbar} x^{2} ]$. We write the quantum action 
$\tilde{S}=\int dt ~ \frac{\tilde{m}}{2} \dot{x}^{2} + \tilde{V}(x)$ with 
$\tilde{V}(x) = \mu x^{2} + \nu/x^{2}$, and choose $\tilde{m}=m$.
The quantum potential has a minimum at $x_{min}=(\nu/\mu)^{1/4}$ and 
$\tilde{v}_{min}=2 \sqrt{\mu \nu}$. 
\begin{figure}
\begin{center}
\epsfig{figure=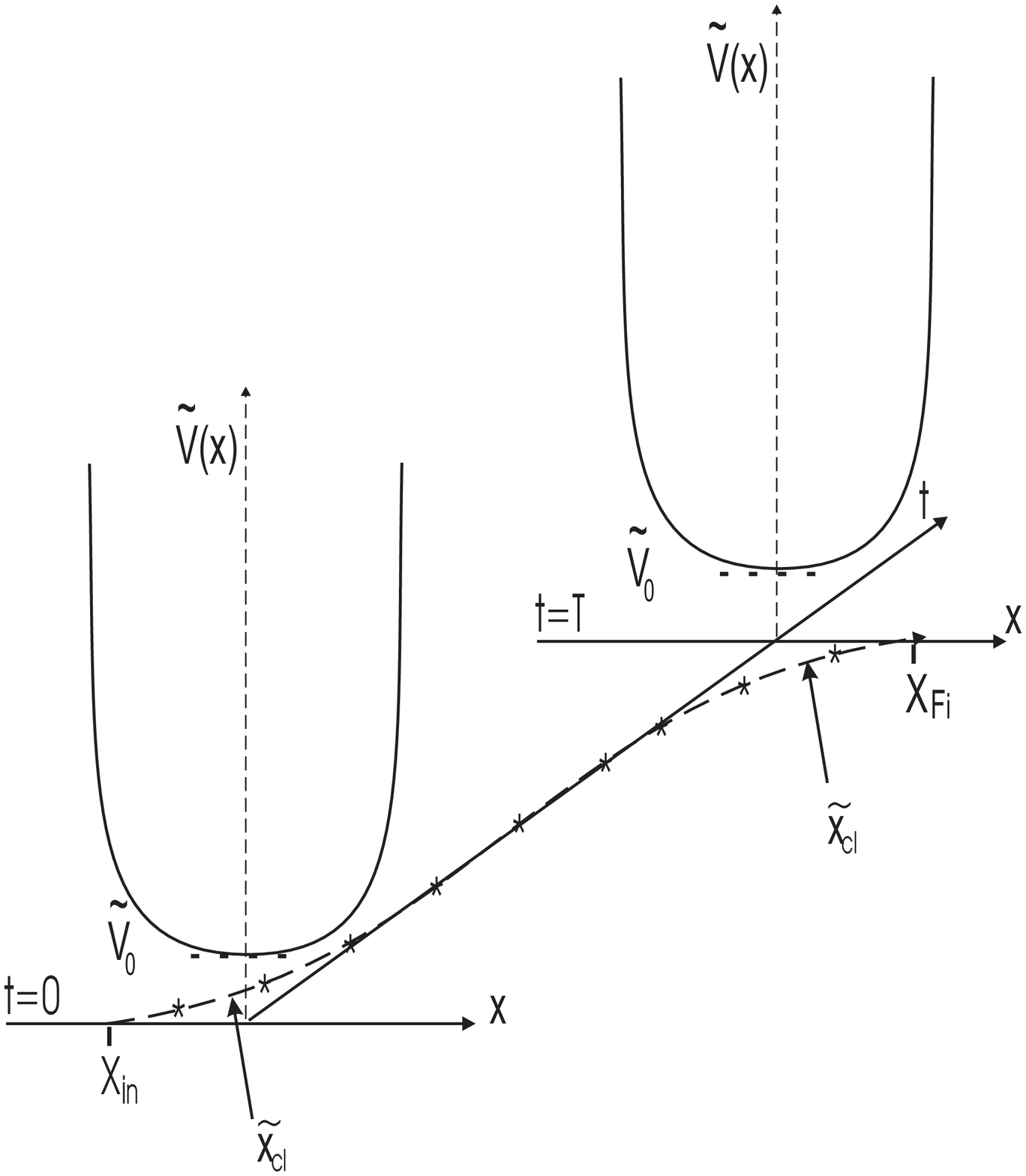,height=5cm,width=.6\linewidth,angle=0}
\end{center}
\noindent
Fig.[1]. Trajectory $\tilde{x}_{cl}$ at bottom-of-the-valley.
\end{figure}
The transformation law with the minimum at $x_{min}$ now reads
\bea
\label{TransformLawMin}
&& 2 m(V(x) - E_{gr}) =  
\nonumber \\
&& 2 \tilde{m}(\tilde{V}(x) - \tilde{v}_{min}) 
- \frac{\hbar}{2} \frac{ \frac{d}{dx} 2 \tilde{m} (\tilde{V}(x) - \tilde{v}_{min})}
{ \sqrt{2 \tilde{m}( \tilde{V}(x) - \tilde{v}_{min} ) } } 
\mbox{sgn}(x-x_{min}) ~ .
\eea
This determines $\mu=\frac{1}{2} m \omega^{2}$, $\nu=\frac{\hbar^{2}}{2 m}[1/2 + \gamma]^{2}$ and reproduces the exactly $E_{gr}$. The wave function derived from the quantum action is now given by (for $x > x_{min}$)
\be
\label{GroundStateLawMin}
\psi_{gr}(x) = \frac{1}{N} ~ e^{ - \int_{x_{min}}^{x} dx' ~ 
\sqrt{2 \tilde{m}( \tilde{V}(x') - \tilde{v}_{min} ) }/\hbar }, \ee
which reproduces the exact wave function. Moreover, one finds that 
$E_{gr}=\tilde{v}_{min}$ and the location of the minimum of the quantum potential coincides exactly with the maximum of the wave function. 
This is an example where the renormalized action has the same structure as the classical action, and only a single coefficient ($1/x^{2}$) has been tuned.

The previous results can be generalized to 3-D. Consider a rotationally invariant potential $V(r)$ with a unique minimum at $r=0$ and giving an s-wave 
ground state, $\psi_{gr}(\vec{x}) = Y_{00}(\theta, \phi) \phi_{gr}(r)$. The path $\tilde{x}_{cl}$ (bottom-of-the-valley) minimizes the action.  
It starts out from $\vec{x}_{in}$ goes in radial direction towards the origin, and after a long time departs from the origin in radial direction to arrive at $\vec{x}_{fi}$. This gives, in analogy to Eq.(\ref{ActFeynKac}), 
\bea
\label{ActFeynKac3D}
\tilde{S}[\tilde{x}_{cl}]|_{x_{in},0}^{x_{fi},T} 
= \tilde{v}_{0} T + \left( 
\int_{0}^{r_{in}} + \int_{0}^{r_{fi}} dr ~  
\sqrt{2 \tilde{m}(\tilde{V}(r) -\tilde{v}_{0}) } 
\right) ~ .
\eea
The ground state properties are given by
\be
\label{GroundStateLaw3D}
\phi_{gr}(r) = \frac{1}{N} ~ e^{ - \int_{0}^{r} dr' ~ 
\sqrt{2 \tilde{m}( \tilde{V}(r') - \tilde{v}_{0} ) }/\hbar }, ~~~ 
E_{gr} = \tilde{v}_{0} ~ .
\ee
Finally, the transformation law reads  
\bea
\label{TransformLaw3D}
&& 2 m(V(r) - E_{gr}) =  2 \tilde{m}(\tilde{V}(r) - \tilde{v}_{0}) 
\nonumber \\
&& - \frac{\hbar}{2} \frac{ \frac{d}{dr} 2 \tilde{m} (\tilde{V}(r) - \tilde{v}_{0})}
{ \sqrt{2 \tilde{m}( \tilde{V}(r) - \tilde{v}_{0} ) } }  
- \frac{2 \hbar}{r} \sqrt{ 2 \tilde{m}( \tilde{V}(r) - \tilde{v}_{0} ) } ~ .
\eea
Eaxamples. (a) For the harmonic oscillator (mass $m$ and potential 
$V(r) = \frac{1}{2} m \omega^{2} r^{2}$),
choosing $\tilde{m}=m$, Eq.(\ref{TransformLaw3D}) 
determines the quantum potential (coinciding with the classical one up to an additive constant) and Eq.(\ref{GroundStateLaw3D})
yields the exact ground state wave function 
$\phi_{gr}(r) = \frac{1}{N} \exp[ - m \omega r^{2}/(2\hbar) ]$, 
and energy $E_{gr} = \frac{3}{2} \hbar \omega$. 
  
(b) Secondly, consider the system with mass $m$ and potential 
$V(r) = \tilde{v}_{6} r^{6} - 5 \hbar \sqrt{\tilde{v}_{6}/(2m)} r^{2}$.
Again, Eqs.(\ref{TransformLaw3D},\ref{GroundStateLaw3D}) 
determine the quantum potential and reproduce the 
exact ground state wave function 
$\phi_{gr}(r) = \frac{1}{N} \exp[ - \sqrt{2 m \tilde{v}_{6}} r^{4}/(4\hbar) ]$
and energy $E_{gr}= 0$.

(c) Finally, let us consider the radial motion (stationary Schr\"odinger equation) in a fixed sector of angular momentum $l>0$ in the hydrogen atom.  
Classically, this corresponds to $S=\int dt T + V$, $T=\frac{m}{2} \dot{r}^{2}$, $V_{l}(r) = \frac{\hbar^{2}l(l+1)}{2m}/r^{2} - e^{2}/r$ (centrifugal plus Coulomb term).
In the zero temperature limit, the transition amplitude is projected onto the states of lowest energy compatible with quantum number $l$.
This is the state with quantum number $n=l+1$ (held fixed in the following).
The radial Schr\"odinger equation gives the energy $E_{n} = R/n^{2}$ 
($R$ Rydberg constant) and the wave function 
$\Phi_{n,l}(r) = 1/N_{n,l} (r/a_{0})^{l} \exp[-r/(n a_{0})]$ 
($a_{0}$ Bohr radius). For the quantum action we make the ansatz $\tilde{m}=m$ and $\tilde{V}(r) = \mu/r^{2} -\nu/r$. 
The quantum potential has a minimum at $r_{min}= \mu/\nu$, with $\tilde{v}_{min}=-\nu^{2}/(4\mu)$.
The transformation law reads now
\bea
\label{TransformLaw3DMin}
&& 2 m(V(r) - E_{n}) =  2 \tilde{m}(\tilde{V}(r) - \tilde{v}_{min}) 
\nonumber \\
&& - \frac{\hbar}{2} 
\frac{ \frac{d}{dr} 2 \tilde{m} (\tilde{V}(r) - \tilde{v}_{min})}
{ \sqrt{2 \tilde{m}( \tilde{V}(r) - \tilde{v}_{min} ) } } sgn(r -r_{min}) 
\nonumber \\
&& - \frac{2 \hbar}{r} 
\sqrt{ 2 \tilde{m}( \tilde{V}(r) - \tilde{v}_{min} ) } sgn(r - r_{min}) ~ .
\eea
From this we determine the coefficients of the quantum action $\mu = \frac{\hbar^{2}}{2m} l^{2}$, $\nu = e^{2} l/(l-1)$ and reproduce exactly the energy $E_{n}$. The quantum action also predicts the wave function, given by (for $r > r_{min}$) 
\be
\label{GroundStateLaw3DMin}
\Phi_{gr}(r) = \frac{1}{N} ~ e^{ - \int_{r_{min}}^{r} dr' ~ 
\sqrt{2 \tilde{m}( \tilde{V}(r') - \tilde{v}_{min} ) }/\hbar } ~ ,
\ee
which coincides with the exact wave function for $n=l+1$. Also we find that 
$E_{n}$ coincides with the minimum of the quantum potential, and the location of this minimum coincides with the maximum of the radial wave function.
The renormalized action has the same structure as the classical action, but both the centrifugal and the Coulomb term get tuned.
It is remarkable that the quantum potential has only terms with exactly the same $r$-dependence, which is a necessary condition for the existence of the Runge-Lenz vector and at the quantum level of the higher $O(4)$ symmetry. 
In this case the quantum action at zero temperature is capable of reproducing a part of the excitation spectrum. 
Finally, the previous results valid for $l>0$ can be 
continued to $n=1, ~ l=0$ and reproduce the exact ground state energy and wave function.
\begin{figure}
\begin{center}
\epsfig{figure=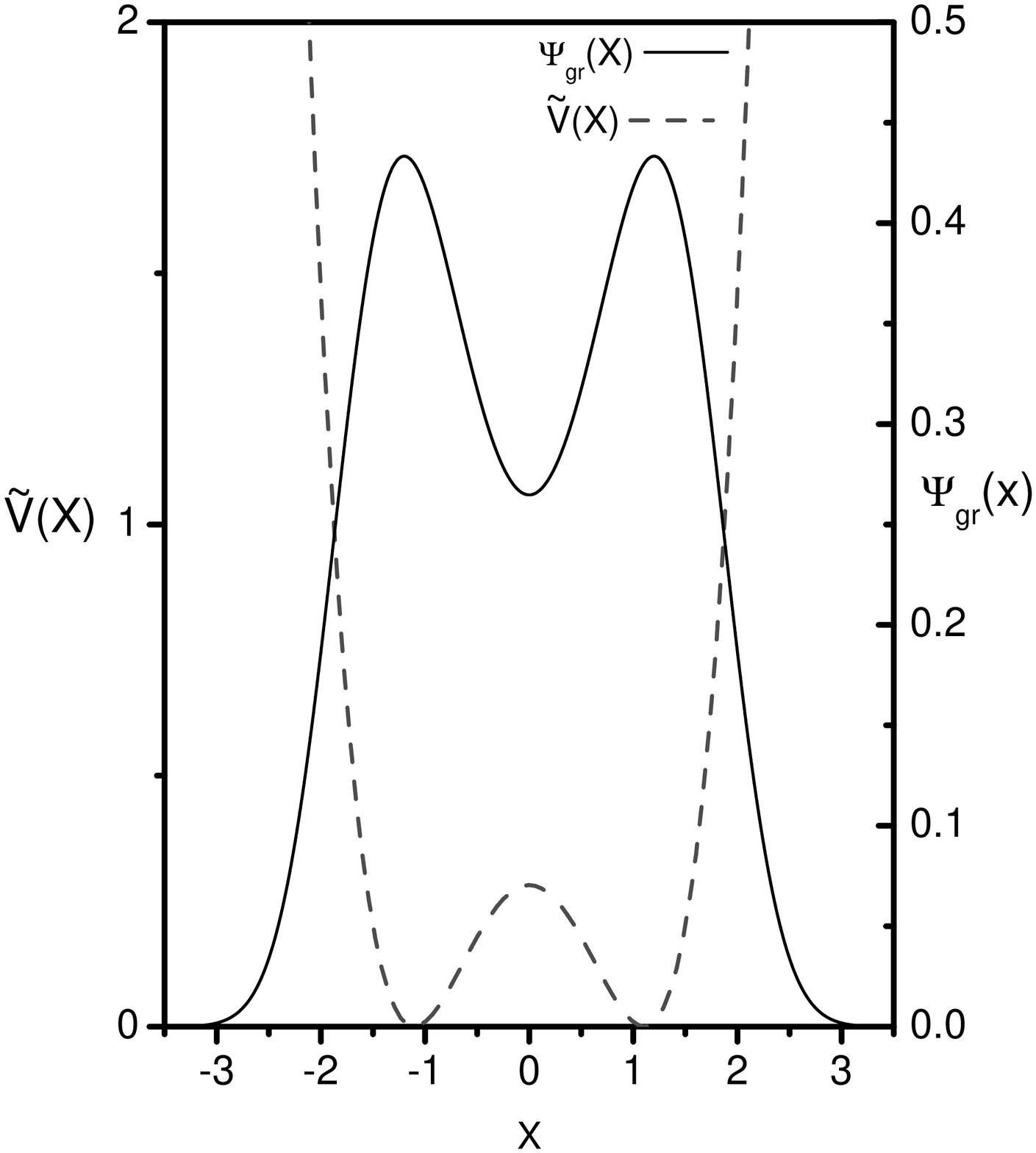,height=7cm,width=.9\linewidth,angle=0}
\end{center}
\noindent
Fig.[2]. Double-well quantum potential and bi-stable ground state.
\end{figure}
We found at the zero temperature
(i) analytic expressions for the quantum action, ground state wave function and energy. The transformation law does not specify both, potential and mass, but only the product $\tilde{m} (\tilde{V} - \tilde{v}_{0})$. The reason is that this product is an invariant  
under the global scale transformation ($\alpha > 0$)
$\tilde{m} \to \tilde{m}/\alpha$, 
$\tilde{V}(x) \to \alpha \tilde{V}(x)$,  
$T \to T/\alpha$  
(similar to a global gauge transformation in gauge theory).
Another invariant is $\tilde{S}[\tilde{x}_{cl}]$.
Applying the same scale transformation to $m$, $V(x)$ and $T$
leaves invariant the Q.M. transition amplitude $G(y,T;x,0)$ and the 
wave function of the ground state and excited states. 
(ii) While the quantum action is local,
the standard effective action displays non-localities, as discussed by 
Gosselin et al. \cite{QM_Renorm}. This shows up in the kinetic term given by an asymptotic series of increasing order time derivatives \cite{Cametti:99}.

{\it 3. Quantum action for double-well potential}.-- 
The scalar $\phi^{4}$ theory in $QFT$ has a symmetric and a spontaneously broken phase, corresponding to a degenerate vacuum. The $QM$ analogue of this is the  symmetric double-well potential. The $QM$ analogue of the symmetric phase is a ground state wave function with a single maximum, while the degenerate vacuum has its analogue in a ground state with two maxima. This distinction 
visible in the wave function also shows up in the quantum action, i.e. in the quantum potential. One can show that the location of the maxima of the wave function coincide with the minima of the quantum potential. As an example, 
we considered $m=1$, $V(x) = \frac{1}{2} -2 x^{2} + \frac{1}{2} x^{4}$. 
We present (Fig.[2]) the ground state computed from the stationary Schr\"{o}dinger equation and the double-well quantum potential, obtained by fitting the quantum action to transition amplitudes.

{\it 4. Renormalisation group equation for temperature dependence of action parameters}.-- In $QFT$ the dependence of action parameters upon variation of a scale parameter (cut-off $\Lambda$, lattice spacing $a_{s}$ and $a_{t}$) is governed by a renormalisation group equation (Callan-Symanzik).
In the $QM$ system considered here, we are close to the 
continuum limit ($\Delta x/a_{B} = 0.045$ and $\Delta t/a_{B}=5 \times 10^{-6}$, $a_{B}$ is the groundstate Bohr radius of the system considered below (Tab.[1])). 
As scale and temperature dependence of the action are similar, we apply 
here the term renormalisation group to describe temperature dependence. We consider the transition amplitude as a function of 
$x$ and $t$, keeping initial data $x_{in}$, $t_{in}$ fixed. It satisfies the Schr\"odinger equation and initial condition
\bea
\label{SchrodEq}
- \hbar \frac{d}{d t} G(x,t;x_{in},t_{in})  
&=& \left[ - \frac{\hbar^{2}}{2 m} 
\frac{d^{2}}{dx^{2}} + V(x) \right] G(x,t;x_{in},t_{in}) 
\nonumber \\
\lim_{t \to t_{in}} G(x,t;x_{in},t_{in}) &=& \delta(x - x_{in}) ~ .
\eea
In the limit $t \to t_{in}$, the transition amplitude 
is given by the classical action, Eq.(\ref{DiracAction}),
consistent with the initial condition, Eq.(\ref{SchrodEq}). 
Going over to inverse temperature $\beta$, the conjecture
$G_{ij} = \tilde{Z}_{\beta} \exp[ - \tilde{\Sigma}_{\beta,ij} ]$ and 
Eq.(\ref{SchrodEq}) imply
\be
\label{SchrodEqSigma}
- \frac{1}{\tilde{Z}^{\beta}} \frac{d \tilde{Z}_{\beta}}{d \beta} 
+ \frac{d \tilde{\Sigma}_{\beta} }{ d\beta } + \frac{\hbar^{2}}{2m} 
[ (\frac{d \tilde{\Sigma}_{\beta}}{dx})^{2} - \frac{d^{2} \tilde{\Sigma}_{\beta}}{dx^{2}} ] 
- V = 0 .
\ee
$\tilde{\Sigma}_{\beta}$ is given by the quantum action along its classical trajectory from $x_{in}, \beta_{in}=0$ to $x, \beta$
\bea
\label{SigmaParam}
\tilde{\Sigma}_{\beta} &=& \tilde{S}_{\beta}[\tilde{x}_{cl}]|_{x_{in},0}^{x,\beta}
= \int_{x_{in},0}^{x,\beta} d\beta' ~ \frac{\tilde{m}}{2 \hbar^{2}} 
(\frac{ d \tilde{x}_{cl}}{d \beta'})^{2} + \tilde{V}(\tilde{x}_{cl}) 
\nonumber \\
&=&  \tilde{\Sigma}_{\beta} \left[ \tilde{m}(\beta),\tilde{v}_{0}(\beta),\tilde{v}_{1}(\beta),\dots,
x,\beta \right] ~ . 
\eea 
It is convenient to parametrize the potential $\tilde{V}(x)$ 
in terms of $\tilde{v}_{k} |x|^{k}$, if $V(x)$ is parity symmetric.
In general the number of terms is infinite. The weight of higher terms decreases rapidly (Tab.[1]). Eqs.(\ref{SigmaParam},\ref{SchrodEqSigma}) yield the  renormalisation group equation for the action parameters, 
$\tilde{m}(\beta), \tilde{v}_{0}(\beta),\tilde{v}_{1}(\beta),\dots$
as function of $\beta$,
\bea
\label{RenormGroup}
&& - \frac{1}{\tilde{Z}_{\beta}} \frac{d \tilde{Z}_{\beta}}{d \beta} 
+  
\frac{\partial \tilde{\Sigma}_{\beta}}{\partial \tilde{m}} 
\frac{\partial \tilde{m}}{\partial \beta } +
\sum_{k} \frac{\partial \tilde{\Sigma}_{\beta}}{\partial \tilde{v}_{k}} 
\frac{\partial \tilde{v}_{k}}{\partial \beta } +
\frac{\partial \tilde{\Sigma}_{\beta}}{\partial \beta} 
\nonumber \\
&& + \frac{\hbar^{2}}{2m} [ (\frac{d \tilde{\Sigma}_{\beta}}{dx} )^{2} 
- \frac{d^{2} \tilde{\Sigma}_{\beta}}{dx^{2}} ] - V = 0 .
\eea
This equation is valid for all $x$, $x_{in}$. The parameters $\tilde{m}(\beta)$, $\tilde{v}_{k}(\beta)$ are independent of $x$, $x_{in}$. This constitutes a system of equations to determine 
$\partial \tilde{Z}_{\beta}/\partial \beta$,
$\partial \tilde{m}/\partial \beta$ and $\partial \tilde{v}_{k}/\partial \beta$.
The solution of the differential equations requires 
initial values. 
For $\beta=0$, Eq.(\ref{DiracAction}) suggests to choose as initial values 
$\tilde{Z}_{\beta}(\beta \to 0) \sim Z(\beta \to 0)$ (note: singularity at origin), $\tilde{m}(\beta=0) = m$, 
$\tilde{v}_{k}(\beta=0) = v_{k}, ~ k=0,1,\dots$.
Starting from initial values the renormalisation group equation determines the flow of the renomalized parameters when $\beta$ increases. 

For the system $m=1$, $V(x) = x^{2} + 0.01 x^{4}$, we have carried out a numerical solution of the flow of the quantum action parameters, by solving a set of differential equations, Eq.(\ref{RenormGroup}), for a common initial point and a number of end points. We have compared this with a global fit of the quantum action to $QM$ transition amplitudes (see Ref.\cite{Jirari1}). The comparison is shown in Fig.[3]. One observes reasonable agreement in the range of $\beta$ between 0.5 and 4.

\begin{figure}
\begin{center}
\epsfig{figure=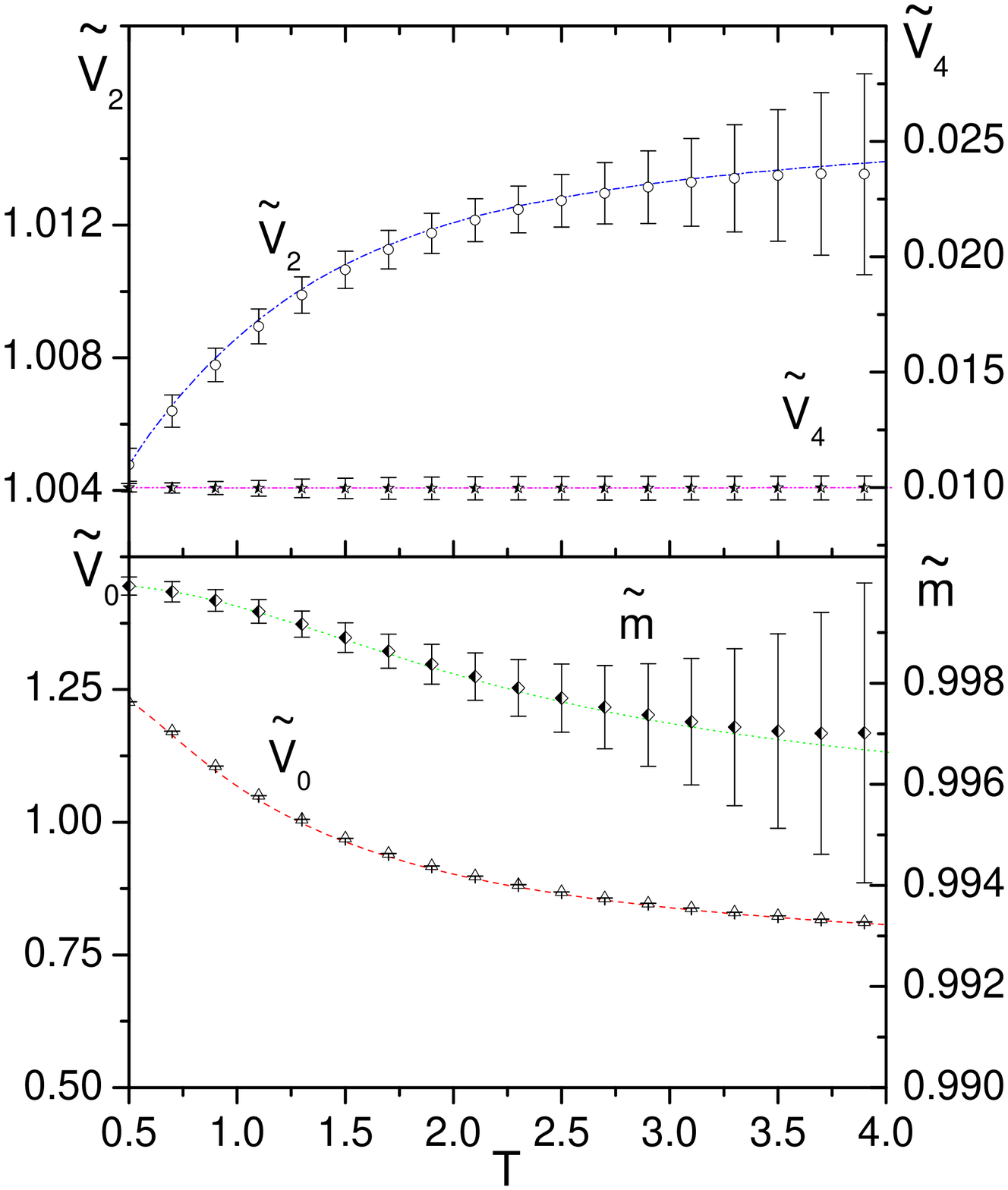,height=8cm,width=1.0\linewidth,angle=0}
\end{center}
\noindent
Fig.[3]. Quantum action parameters. Comparison flow equation vs. fit to transition amplitudes.
\end{figure}

\bigskip

{\it Acknowledgements}.-- H.K. and K.M. are grateful for support by NSERC Canada. X.Q. Luo has been supported by NNSF of China.
H.K. is grateful for discussions with L.S. Schulman.

\end{multicols}


\begin{thebibliography}{20}

\bibitem{QM_Renorm} K.S.~Gupta and S.G.~Rajeev, Phys. Rev. D48(1993)5940; 
C.~Manuel and R.~Tarrach, Phys. Lett. B328(1994)113; J.~Polonyi, Ann. Phys. 252(1996)300; P.~Gosselin and H.~Mohrbach, J. Phys. A33(2000)6343. 

\bibitem{Jirari1} H.~Jirari, H.~Kr\"{o}ger, X.Q.~Luo, K.J.M.~Moriarty and S.G.~Rubin, Phys. Rev. Lett. 86(2001)187.

\bibitem{Milner:01} V.~Milner, J.L.~Hansen, W.C.~Campbell and \\
M.G.~Raizen, 
Phys. Rev. Lett. 86(2001)1514.

\bibitem{Friedman:01} N.~Friedman, A.~Kaplan, D.~Carasso and N.~Davidson, Phys. Rev. Lett. 86(2001)1518.

\bibitem{Dembrowski:01} C.~Dembrowski, H.D.~Gr\"{a}ff, A.~Heine, T.~Hesse, H.~Rehfeld and A.~Richter, Phys. Rev. Lett. 86(2001)3284.
 
\bibitem{Jirari2} H.~Jirari, H.~Kr\"{o}ger, X.Q.~Luo, K.J.M.~Moriarty and S.G.~Rubin, Phys. Lett. A281(2001)1.

\bibitem{WCNA} H.~Jirari, H.~Kr\"{o}ger, X.Q.~Luo, and K.J.M.~Moriarty, 
Proc. WCNA, Catania (2000), Nonlin. Analys. 47/48(2001)473-484, quant-ph/0102032.

\bibitem{Cametti:99} F.~Cametti, G.~Jona-Lasinio, C.~Presilla and 
F.~Toninelli, quant-ph/9910065, 
Proc. School of Physics Enrico Fermi, IOS Press, Amsterdam (2000), p.431.

\bibitem{Schulman} L.S. Schulman, {\sl Techniques and Applications of 
Path Integration}, J. Wiley\& Sons, New York (1981).

\end{thebibliography}
\end{document}